\documentclass[aps,prb,amsmath,twocolumn,amssymb,floatfixng,showpacs,
superscriptaddress,footinbib]{revtex4-1}
\pdfoutput=1
\usepackage[dvips]{graphics}
\usepackage{bm}
\usepackage{epsfig}
\usepackage{enumerate}
\usepackage{subfigure}
\usepackage{color}
\usepackage{graphicx}
\usepackage{hyperref}
\hypersetup{
    pdfnewwindow=true,      
    colorlinks=true,       
    linkcolor=magenta,          
    citecolor=blue,        
    filecolor=blue,      
    urlcolor=blue        
}

\newcommand{\etal}{{\it et al.~}}
\newcommand{\ie}{{\it i.e., }}

\begin{document} 
\title{Signature of tilted Dirac cones in Weiss oscillations of $8-Pmmn$ borophene} 

\author{SK Firoz Islam}
\email{firoz@iopb.res.in}
\author{A. M. Jayannavar}
\email{jayan@iopb.res.in}

\affiliation{Institute of Physics, Sachivalaya Marg, Bhubaneswar-751005, India}
\begin{abstract}
Polymorph of $8-Pmmn$ borophene exhibits anisotropic tilted Dirac cones. In this work, we explore the consequences
of the tilted Dirac cones in magnetotransport properties of a periodically modulated borophene.
We evaluate modulation induced diffusive conductivity by using linear response theory in low temperature regime.
The application of weak modulation (electric/magnetic or both) gives rise to the magnetic field dependent
non-zero oscillatory drift velocity which causes Weiss oscillation in the longitudinal conductivity at low magnetic field.
The Weiss oscillation is studied in presence of an weak spatial electric, magnetic and both modulations individually.
The tilting of the Dirac cones gives rise to additional contribution to the Weiss oscillation in longitudinal conductivity.
Moreover, it also enhances the frequency of the Weiss oscillation and modifies its amplitude too.
Most remarkably, It is found that the presence of out-of phase both \ie electric and magnetic modulations can cause
a sizable valley polarization in diffusive conductivity. The origin of valley polarization lies in the opposite tilting of the
two Dirac cones at two valleys.
\end{abstract}
\maketitle
\section{Introduction}\label{intro}
In recent times, Dirac materials have attracted intense research interests after the most celebrated discovery of atomically thin
two dimensional (2D) hexagonal carbon allotrope-graphene\cite{RevModPhys.81.109,sarma2011electronic}, owing to their peculiar
band structure and applications in next generation of nanoelectronics. The polymorph of borophene with tilted anisotropic
Dirac cones (named as $8-Pmmn$ borophene)\cite{PhysRevLett.112.085502} is the latest 2D material to the family
of Dirac systems. Very recently, experimental confirmation of such material has been reported\cite{PhysRevLett.118.096401}, 
followed by a detail analysis of its ab-initio properties\cite{PhysRevB.93.241405}. Similar to the strained graphene\cite{PhysRevLett.103.046801},
a pseudo magnetic field has been recently predicted in $8-Pmmn$ borophene by using tight-binding model\cite{PhysRevB.94.165403}.
An effective low energy Hamiltonian in the vicinity of Dirac points has been proposed based on symmetry consideration\cite{PhysRevB.94.165403},
which has recently been used to investigate collective excitations (plasmons) \cite{PhysRevB.96.035410}
and optical properties\cite{PhysRevB.96.155418} theoretically.

Magnetotransport properties have always been appreciated for providing a powerful and experimentally reliable tool to probe a 2D
fermionic system. The presence of magnetic field, normal to the plane of the 2D sheet of electronic system, discretizes the energy
spectrum by forming Landau levels (LLs) which manifests itself via oscillatory longitudinal conductivity with inverse magnetic
field-known as Shubnikov-de Hass (SdH) oscillation\cite{feng2005introduction,imry1997introduction}. 
In addition to the SdH oscillation, another type of quantum oscillations appears in low magnetic field regime when the 
$2$D fermionic system is subjected to an weak spatial electric/magnetic modulation. This oscillation is known as Weiss
oscillation which was first observed in magneto-resistance measurements in the electrically modulated  usual
two dimensional electron gas ($2$DEG)~\cite{weiss1989,PhysRevLett.62.1173,PhysRevLett.62.1177}.
The Weiss oscillation is also known as Commensurability oscillation as it is caused by the commensurability of the two length scales
\ie cyclotron orbit's radius near the Fermi energy and the modulation period~\cite{PhysRevLett.63.2120,PhysRevB.41.12850,PhysRevB.46.4667}.
An alternative explanation was also given by Beenakker~\cite{PhysRevLett.62.2020} by using the concept of {\it guiding-center-drift resonance}
between the periodic cyclotron orbit motion and the oscillating drift of the orbit center induced by the potential grating.

Apart from the electric modulation case, magnetic modulation has also been considered 
theoretically~\cite{PhysRevB.47.1466,peeters_super,li1996electrical,PhysRevB.62.91,PhysRevB.81.115308,PhysRevB.45.5986,papp2004giant}
as well as experimentally~\cite{izawa1995,PhysRevLett.74.3009,PhysRevLett.74.3013}. Weiss oscillation has been studied
in Rashba spin-orbit coupled electrically/magnetically modulated $2$DEG and beating pattern was predicted~\cite{PhysRevB.71.125301,islam2012magnetotransport}.
The higher Fermi velocity associated with the linear band dispersion significantly enhances the Weiss oscillation in an electrically
modulated graphene~\cite{PhysRevB.75.125429,nasir2010magnetotransport}. Concurrently, same has been studied in a magnetically
modulated graphene too and enhancement of the amplitude and opposite phase in comparison to the case of electrically 
modulated graphene was observed~\cite{PhysRevB.77.195421} . Similar investigations have been carried out in electrically
modulated bilayer graphene~\cite{PhysRevB.85.245426}, silicene~\cite{islam2014beating,PhysRevB.90.125444}, $\alpha-\mathcal{T}_3$
-lattice~\cite{PhysRevB.96.045418} and phosphorene~\cite{Tahir_JPCM}. However, magnetotransport properties of
modulated borophene are yet to be explored.

In this work, we investigate the modulation induced longitudinal conductivity of borophene in low temperature regime by using
the linear response theory. First, we obtain exact LLs and corresponding density of states (DOS) in $8$-$Pmmn$ borophene.
Numerically, we notice that the tilting of the Dirac cones lowers the Fermi level. We observe modulation induced Weiss oscillation
in the longitudinal conductivity in low magnetic field regime. Interestingly, we find that the opposite tilting of the 
Dirac cones at two valleys can cause sizable valley polarization in the longitudinal conductivity at low magnetic field regime under the
combined effects of out-of phase electric and magnetic modulation, which is in contrast to the non-tilted isotropic Dirac cones-graphene.
Moreover, the tilting of the Dirac cones also enhances the frequency of Weiss oscillation.

The paper is organized as follows. In Sec.~\ref{sec2}, we introduce the low energy effective Hamiltonian and
derive LLs. The effect of tilting of Dirac cones on the Fermi level and DOS are also included in this section.
The Sec.~\ref{sec3} devoted to calculate the modulation induced Weiss oscillation in the longitudinal conductivity.
Finally, we summarize and conclude in Sec.~\ref{sec4}.

\section{Model Hamiltonian and Landau level formation}\label{sec2}
We start with the single particle low energy effective model Hamiltonian of the tilted anisotropic Dirac cones
as\cite{PhysRevB.94.165403,PhysRevB.96.035410}
\begin{equation}
 H=\xi(v_xp_x\sigma_x+v_yp_y\sigma_y+v_tp_y\sigma_0),
\end{equation}
where the first two terms corresponds to the kinetic energy term and the last term describes the tilted nature of Dirac cones. 
The two Dirac points are at ${\bf k=\pm k_D}$, described by the valley index $\xi=\pm$. Hereafter, we shall denote two valleys
as $K$ and $K'$, corresponding to $\xi=+$ and $\xi=-$, respectively. Three velocities are given by $\{v_x,v_y,v_t\}=\{0.86,0.69,0.32\}$
in units of $v_0=10^{6}$ m/s. The velocity $v_t$ arises due to the tilting of the Dirac cones.
Also, $(\sigma_x,\sigma_y)$ are the Pauli matrices and $\sigma_0$  is identity matrix. The energy dispersion of the above
Hamiltonian can be readily obtained as
 \begin{equation}\label{band}
  E_{\lambda,k}^{\xi}=\xi\hbar v_tk_y+\lambda\hbar\sqrt{v_x^2k_x^2+v_y^2k_y^2},
 \end{equation}
where $\lambda=\pm$ is the band index and the $2$D momentum vector is given by ${\bf k}=\{k_x,k_y\}$.
\begin{figure}[!thpb]
\centering
\includegraphics[height=5cm,width=0.80 \linewidth]{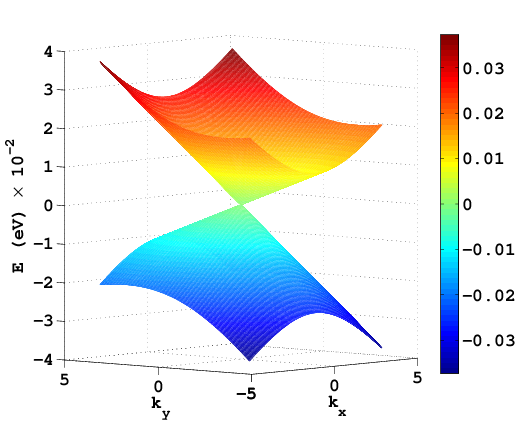}
\caption{(Color online) The energy band dispersion in k-space representing Eq.~(\ref{band}). The momentum vectors are normalized
by $k_0=10^{7}$/m.}
\label{band}
\end{figure}
The energy dispersion for $K$-valley is shown in Fig.~(\ref{band}), which is tilted along $k_y$ due to the presence of $v_t$ term. 
In $K'$-valley, dispersion will be identical except the tilting is in opposite direction. In addition to this, Dirac cones are
anisotropic which is in contrast to graphene. Note that because of the tilted feature of the Dirac cones, particle-hole symmetry
is broken in borophene.
\subsubsection{Inclusion of magnetic field}
The perpendicular magnetic field (${\bf B}=B\hat{z}$) is incorporated via the Landau-Peierls substitution as: ${\bf p}\rightarrow {\bf p+eA}$ in
low energy single electron effective Hamiltonian of borophene, lying in the $x$-$y$ plane, as
\begin{equation}
  \mathcal{H}= \xi[v_xp_x\sigma_x+v_y(p_y+eBx)\sigma_y+v_t(p_y+eBx)\sigma_0],
\end{equation}
under the Landau gauge ${\bf A}=(0,xB,0)$. Here, $A$ is the magnetic vector potential. The commutator relation $[\mathcal{H},p_y]=0$ guarantees
the free particle nature of electron along the $y$-direction. Using this fact, the above Hamiltonian reduces to
\begin{equation}\label{hamil}
\mathcal{H}=\xi\left\{\frac{\hbar v_t}{l_c}X\sigma_0+\frac{\hbar v_c}{l_c}\left[\sqrt{\frac{v_x}{v_y}}\sigma_xP+\sqrt{\frac{v_y}{v_x}}\sigma_y X\right]\right\}, 
\end{equation}
where $l_c=\sqrt{\hbar/eB}$ is the magnetic length, $P=-i\partial/\partial(x/l_c)$, $v_c=\sqrt{v_xv_y}$ and $X=(x+x_0)/l_c$ with
the center of cyclotron orbit is at $x=-x_0=-k_yl_c^2$. The above Hamiltonian is now similar to the case of monolayer graphene under
crossed electric and magnetic field\cite{lukose2007novel} except the velocity anisotropy inside the third bracket. The first term
is analogous to a pseudo in-plane effective electric field $E_{eff}=\xi\hbar v_t/(el_{c}^2)$. The typical values of the pseudo-electric field
are $(320\times\rm B)$ kV. Now Eq.~(\ref{hamil}) can be re-written as 
\begin{equation}
 \mathcal{H}=eE_{eff}(x+x_0)\sigma_0
 +\xi\hbar \omega_c\left[\begin{array}[c]{c c}0 & -ia\\ia^{\dagger} &0\end{array}\right],
\end{equation}
where $\omega_c (=v_c/l_c)$ is the cyclotron frequency and ladder operators are defined as: $a=(\tilde{X}+i\tilde{P})/\sqrt{2}$
and $a^{\dagger}=(\tilde{X}-i\tilde{P})/\sqrt{2}$. Here, $\tilde{X}=\sqrt{\frac{v_y}{v_x}}X$ and $\tilde{P}=\sqrt{\frac{v_x}{v_y}}P$,
which satisfy the commutator relation $[\tilde{X},\tilde{P}]=i$. In absence of $E_{eff}$, the above Hamiltonian can be diagonalized
to obtain graphene-like LLs
\begin{equation}
 E_{\zeta}=\lambda\hbar\omega_c\sqrt{2n}
\end{equation}
and eigenfunctions as
\begin{equation}\label{wave}
\psi_{\zeta}({\bf r})=\frac{e^{ik_yy}}{\sqrt{2L_y}}\left[\begin{array}[c]{c}
                                   \lambda\phi_{n}(X)\\
                                   -i\xi\phi_{n-1}(X)
                                  \end{array}\right],
\end{equation}
where $\zeta=\{n,k_y\}$ and $\phi_{n}(X)$ is the well known simple harmonic oscillator wave functions. In presence of $E_{eff}$, direct diagonalization
of the above Hamiltonian is difficult. However, there is a standard way, given by V. Lukose \etal in Ref.[\onlinecite{lukose2007novel}],
to solve this problem exactly. An alternative approach of solving this problem in graphene was also  given by NMR Peres \etal \cite{peres2007algebraic}.
Following Ref.[\onlinecite{lukose2007novel}], we transform the above Hamiltonian into a moving frame along the $y$-direction with
velocity $V=E_{eff}/B=v_t$, where the transformed electric field vanishes and magnetic field reduces to $B'=B\sqrt{1-\beta_b^2}$.
Here, $\beta_b=v_t/\sqrt{v_xv_y}(=0.4154)$ is termed as ``tilt parameter''. Note that the role of velocity of light is played by
$v_c$ in borophene whereas in graphene it is $v_F$. In the moving frame, LLs can be obtained as $\tilde{E}_{n,\tilde{ky}}=
\hbar\omega_c\sqrt{2n}(1-\beta_b^2)^{1/4}$. However, required LLs and eigen states in the rest frame can be obtained by
Lorentz boost back transformation as\cite{lukose2007novel,PhysRevB.92.035306}:
\begin{equation}\label{ll}
 E_{\zeta}=\lambda\hbar\omega_c\sqrt{2n}(1-\beta_b^2)^{3/4},
\end{equation}
where the argument of the wave functions becomes 
\begin{equation}
 X'=\frac{(1-\beta_b^2)^{1/4}}{l_c}\left[x+k_yl_c^2+\lambda\frac{\sqrt{2n}l_c\beta_b}{(1-\beta_b^2)^{1/4}}\right]
\end{equation}
after using the Lorentz back transformation of momentum, giving the wave function in rest frame as
\begin{eqnarray}\label{wave}
\Psi_{\zeta}({\bf r})&=&\frac{e^{ik_yy}}{\sqrt{2L_y\gamma}}\Big[\left(\begin{array}[c]{c}
                                  \cosh(\theta/2)\\
                                  -i\sinh(\theta/2)
                                  \end{array}\right)\lambda \phi_{n}(X')\nonumber\\
                                  &&-i\xi\left(\begin{array}[c]{c}
                                  i\sinh(\theta/2)\\
                                  \cosh(\theta/2)
                                  \end{array}\right)\phi_{n-1}(X')\Big]
\end{eqnarray}
with $\tanh\theta=\beta_b$ and $\cosh\theta=\gamma$. Here, we have used the form of hyperbolic rotation matrix as 
\begin{equation}
 e^{-\left(\frac{\theta}{2}\right)\sigma_y}=\left[\begin{array}[c]{c c}
                                  \cosh(\theta/2)&i\sinh(\theta/2)\\
                                  -i\sinh(\theta/2)&\cosh(\theta/2)
                                  \end{array}\right].
\end{equation}
On the other hand, the LLs of graphene under the in-plane real electric field ($E_r$) is given by\cite{lukose2007novel}
\begin{equation}\label{ll_gra}
 E^{g}_{\zeta}=\lambda\hbar\omega_c\sqrt{2n}(1-\beta_g^2)^{3/4}-\hbar k_y\frac{E_r}{B},
\end{equation}
where $\beta_g=\frac{E_r/B}{v_F}$ with $v_F$ is the Fermi velocity. Note that in cyclotron frequency, $v_c$ should be replaced by $v_F$ in graphene.
In Eq.~(\ref{ll}), $\beta_b$ is a constant and acting like a system parameter, whereas $\beta_g$ is tunable and governed by the strength of the
in-plane electric field in graphene. As the tilt parameter ($\beta_b$) is constant, the LLs are protected from being collapsed in borophene
which is in contrast to graphene where LLs may get collapsed under the suitable strength of the electric field (\ie when $\beta_g$ becomes $1$).
Note that the LLs in borophene, derived in Eq.~(\ref{ll}), exhibit $k_y$ degeneracy whereas in graphene [see Eq.~(\ref{ll_gra})]
this degeneracy is removed under the influence of an in-plane electric field. This is because the in-plane electric field in graphene
gives rise to the potential energy as $eEx$ whereas in borophene for pseudo electric field it is $eE_{eff}(x+x_0)$.   
The idea of relativistic Lorentz boost transformation was also used in an organic compound $\alpha-({\rm BEDT-TTF})_{2}I_{3}$
\cite{goerbigEPL}, exhibiting quite similar band structure.

The LLs, derived in Eq.~(\ref{ll}), show that the tilt parameter ($\beta_b$) renormalizes each LLs, which should be reflected in
the longitudinal conductivity oscillations. Before going into conductivity, we now discuss how Fermi energy and DOS are 
affected by the tilting of the Dirac cones.

\subsubsection{Fermi energy and density of states}
In this subsection, we compute the Fermi energy ($E_F$) and DOS in terms of the tilt parameter and the magnetic field. 
In presence of the magnetic field, the Fermi energy can be obtained by solving the following equation self consistently
\begin{equation}\label{fermi}
n_e=\int_{-\infty}^{\infty}D(E)f(E)dE,
\end{equation}
where 
\begin{equation}\label{delta}
D(E)=\frac{g_sg_v}{\Omega}\sum_{\zeta}\delta(E-E_{\zeta})
\end{equation}
\begin{figure}[!thpb]
\centering
\includegraphics[height=6cm,width=0.95 \linewidth]{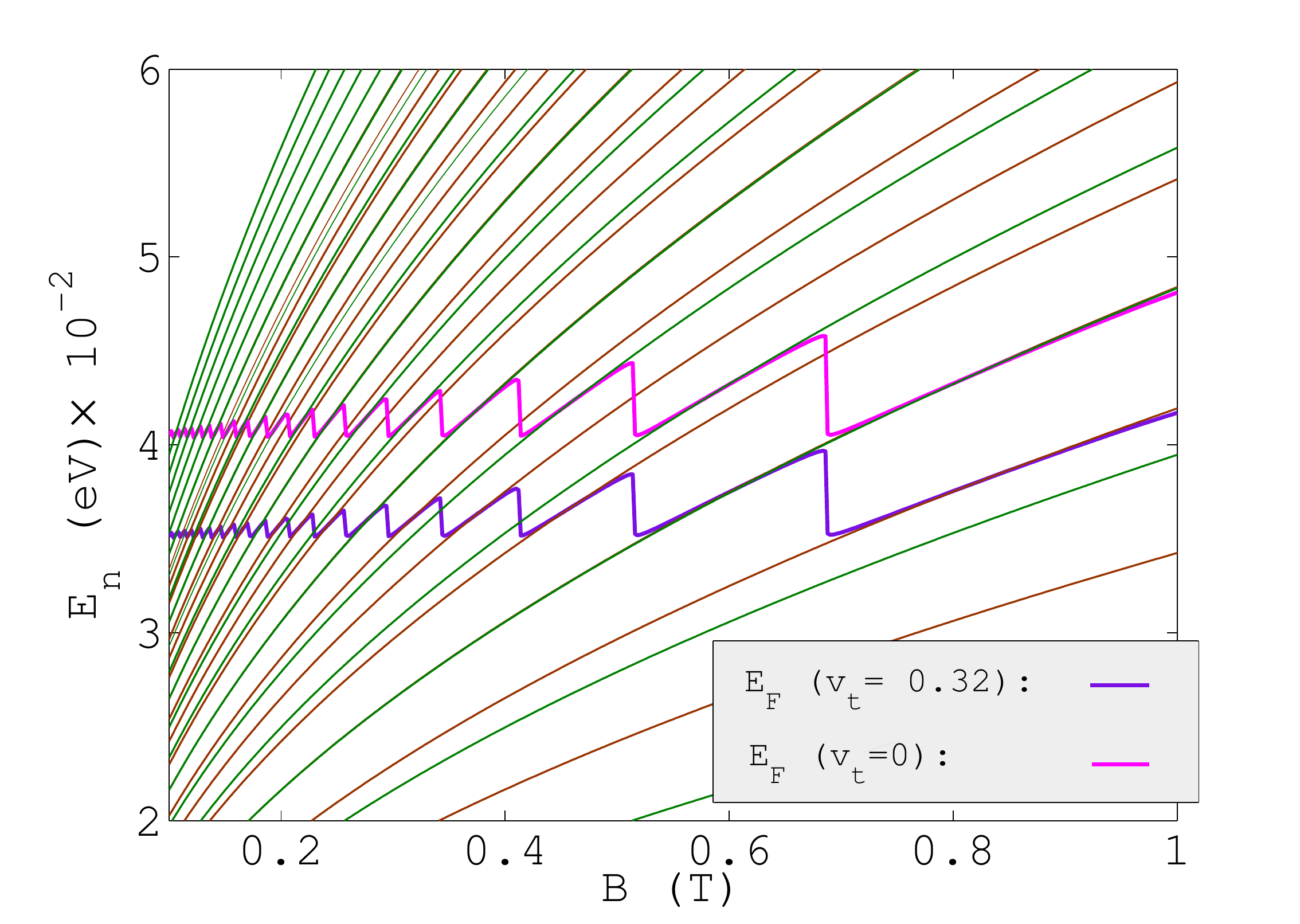}
\caption{(Color online) The behavior of the Fermi energy and the LLs with magnetic field. For the plot of $E_{F}$ versus B,
we use $T=1$ K and carrier density $n_e=10^{15}m^{-2}$. Green and grey solid lines denote first $20$ LLs with and without $v_t$, respectively.}
\label{Fermi}
\end{figure}
is the DOS per unit energy and per unit area. Here, $g_s$ and $g_v$ are the spin and valley degeneracy, respectively. Carrier density 
and the area of the system are denoted by $n_e$ and $\Omega(=L_xL_y)$, respectively.
The Fermi distribution function is given by $f(E)=\left(1+\exp[(E-E_{F})/{k_BT}]\right)^{-1}$. The summation over $k_y$ can be computed by using the fact
that the center of cyclotron orbit is always restricted by the system dimension \ie $0\le |x_0+G_n|\le L_x$ or $0\le k_y\le L_x/l_c^2$.
Then we can replace $\sum_{k_y}\rightarrow \frac{L_y}{2\pi}\int_0^{L_x/l_c^2}dk_y=\Omega/(2\pi l_c^{2})$-known as $k_y$-degeneracy.
The factor $L_y/(2\pi)$ preserves periodic boundary condition. Using these, finally Eq.~(\ref{fermi}) simplifies to
\begin{equation}\label{EF}
\pi n_e l_{c}^2=2\sum_{n}f(E_{n})
\end{equation}
which is solved numerically to plot the Fermi energy as a function of the magnetic field in Fig.~(\ref{Fermi}).
Here we have also substituted spin and valley degeneracy as $g_s=2$ and $g_v=2$, respectively. In the same plot,
first twenty LLs are also shown. The Fermi level is found to be fluctuating between two successive LLs with the
variation of the magnetic field. The amplitude of fluctuation increases with the increase of the magnetic field,
because of the increasing LLs spacing. To understand how the tilting feature of the Dirac cones affects the Fermi energy, we
consider the two cases \ie when $v_t=0$ and $v_t= 0.32$ unit. It can be seen that for the same carrier density, tilt factor ($v_t$)
actually lowers the Fermi level. On the other hand, it causes a shift in the LLs as can be seen from Eq.~(\ref{ll}).
Note that the position of the jumping of Fermi level between two successive Landau levels remains unchanged in both cases.\\
\begin{figure}[thpb]
\centering
{\includegraphics[width=.95\linewidth,height=6cm]{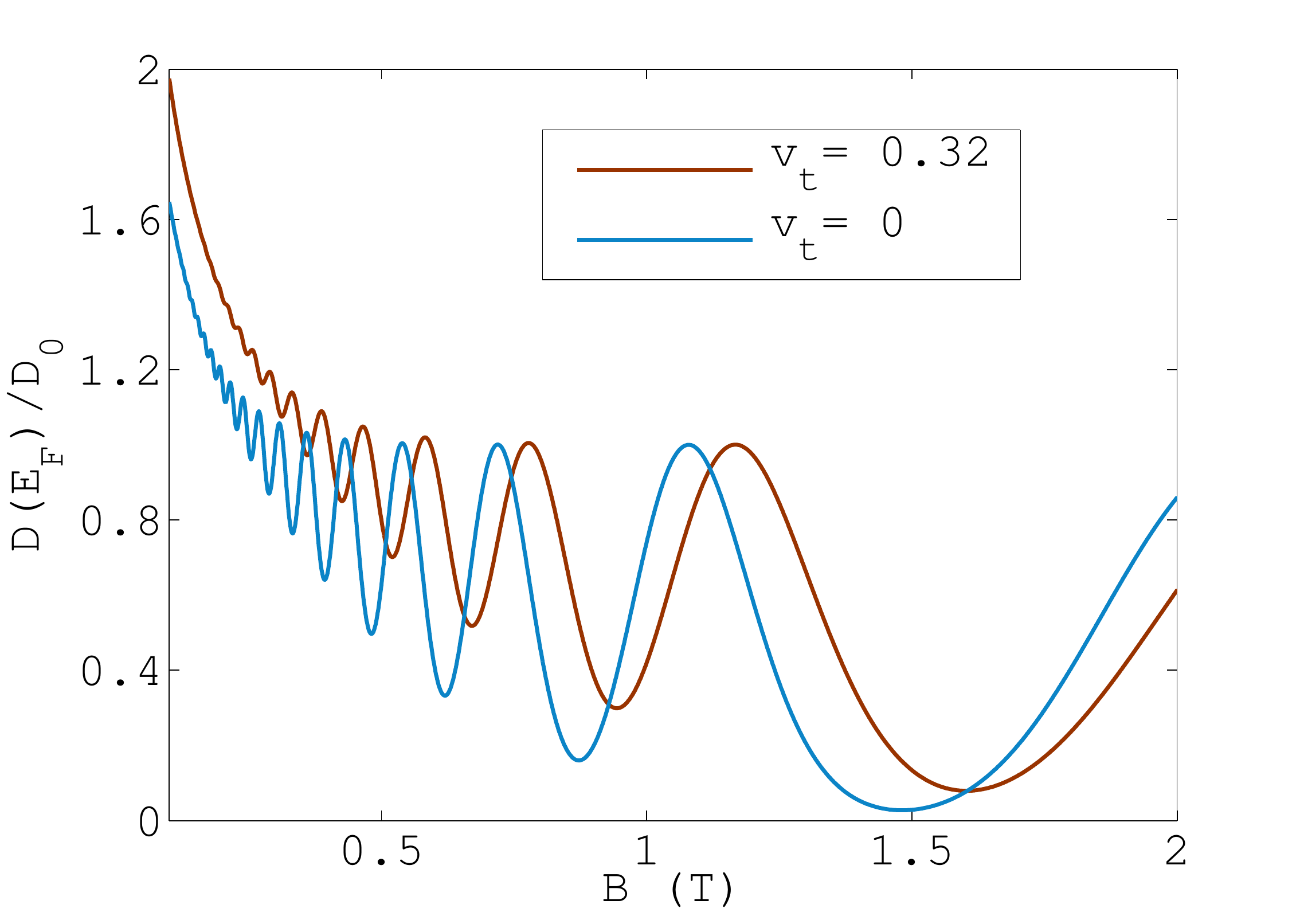}}
\caption{DOS versus magnetic field. The Fermi energy $E_F=0.035$ eV and $0.041$
eV for $v_t=0.32$ and $0$ in units of $v_0$, respectively, as noted from the figure \ref{Fermi}.}
\label{density}
\end{figure}
Now we will examine the effects of the tilting of the Dirac cones on the behavior of the DOS in borophene.
To plot the behavior of DOS, we assume impurity induced Gaussian broadening of
the LLs and hence Eq.~(\ref{delta}) reduces to
\begin{equation}\label{dos}
 D(E)= D_{0}\sum_{n}\exp\left[-\frac{(E-E_n)^2}{2\Gamma_0^2}\right]
\end{equation}
where 
\begin{equation}
D_0=\frac{g_sg_v}{2\pi l_c^2}\frac{1}{\Gamma_0\sqrt{2\pi}}.
\end{equation}
The DOS is plotted in Fig.~(\ref{density}) by using Eq.~(\ref{dos}). It is an established fact\cite{PhysRevB.65.245420,PhysRevB.47.1522}
that the impurity induced LLs broadening in 2D Dirac material is directly proportional to $\sqrt{B}$. To plot dimensionless DOS, we consider LLs
broadening width $\Gamma_0=0.1\hbar\omega_c$. To explore the effects of the tilted Dirac cone, we consider the both cases \ie with and with
out $v_t$. The DOS shows oscillation with the magnetic field-known as SdH oscillation. The presence of $v_t$ is causing a significant impact
on the frequency of the SdH oscillations. It is also observed that below a certain magnetic field, the SdH oscillation vanishes because
of the reduction in the LLs spacing and overlapping of the LLs to each other due to the impurity induced broadening.
\section{Magnetoconductivity}\label{sec3}
In this section, we evaluate magnetoconductivity in presence of a periodic electric/magnetic modulation at low magnetic field regime.
In the presence of a spatial electric/magnetic modulation along the $x$-direction, electron gains a finite
drift velocity along the $y$-direction for which an additional contribution to the $y$ component of the longitudinal conductivity
appears, known as diffusive conductivity\cite{PhysRevB.46.4667}, \ie $\sigma_{yy}=\sigma_{yy}^{\rm dif}+\sigma_{xx}^{\rm col}$. Here, $\sigma_{xx}^{\rm col}$
is the collisional conductivity which arises due to LL induced oscillatory DOS without any external modulation whereas
$\sigma_{yy}^{\rm dif}$ is the diffusive conductivity arises because of the modulation. On the other hand, the longitudinal
conductivity along the $x$-direction is $ \sigma_{xx} = \sigma_{xx}^{\rm col} $ because $\sigma_{xx}^{\rm dif} = 0$.
However, in this work our major focus will be modulation induced diffusive conductivity which
can be evaluated by\cite{charbonneau1982linear,PhysRevB.46.4667}
\begin{equation}\label{diff}
\sigma_{\mu\nu}^{\rm dif}=\frac{\beta e^2}{\Omega}\sum_{\zeta}f_{\zeta}(1-f_{\zeta})\tau(E_\zeta)\mathcal{V}_{\mu}\mathcal{V}_{\nu}
\end{equation}
provided the scattering processes are elastic or quasi elastic. Here, $f_{\zeta}= [1+\exp\{\beta(E_{\zeta}-E_{F})\}]^{-1}$
is the Fermi-Dirac distribution function with $\beta=(k_BT)^{-1}$ where $k_B$ is the Boltzmann constant. In the above formula, $\tau(E_{\zeta})$ denotes
the energy dependent collision time and the group velocity $\mathcal{V}_{\mu(\nu)}=(1/\hbar)\partial E_{\zeta}/\partial k_{\mu(\nu)}$. 
In general, electron does not possess any non-zero drift velocity inside the bulk \ie $\mathcal{V}_x=\mathcal{V}_y=0$.
However, the application of a spatial electric/magnetic modulation can induce a non-zero finite drift velocity
and concurrently gives rise to the diffusive conductivity as discussed below.
\begin{figure*}
 \subfigure[]
 {\includegraphics[width=.49\textwidth,height=5cm]{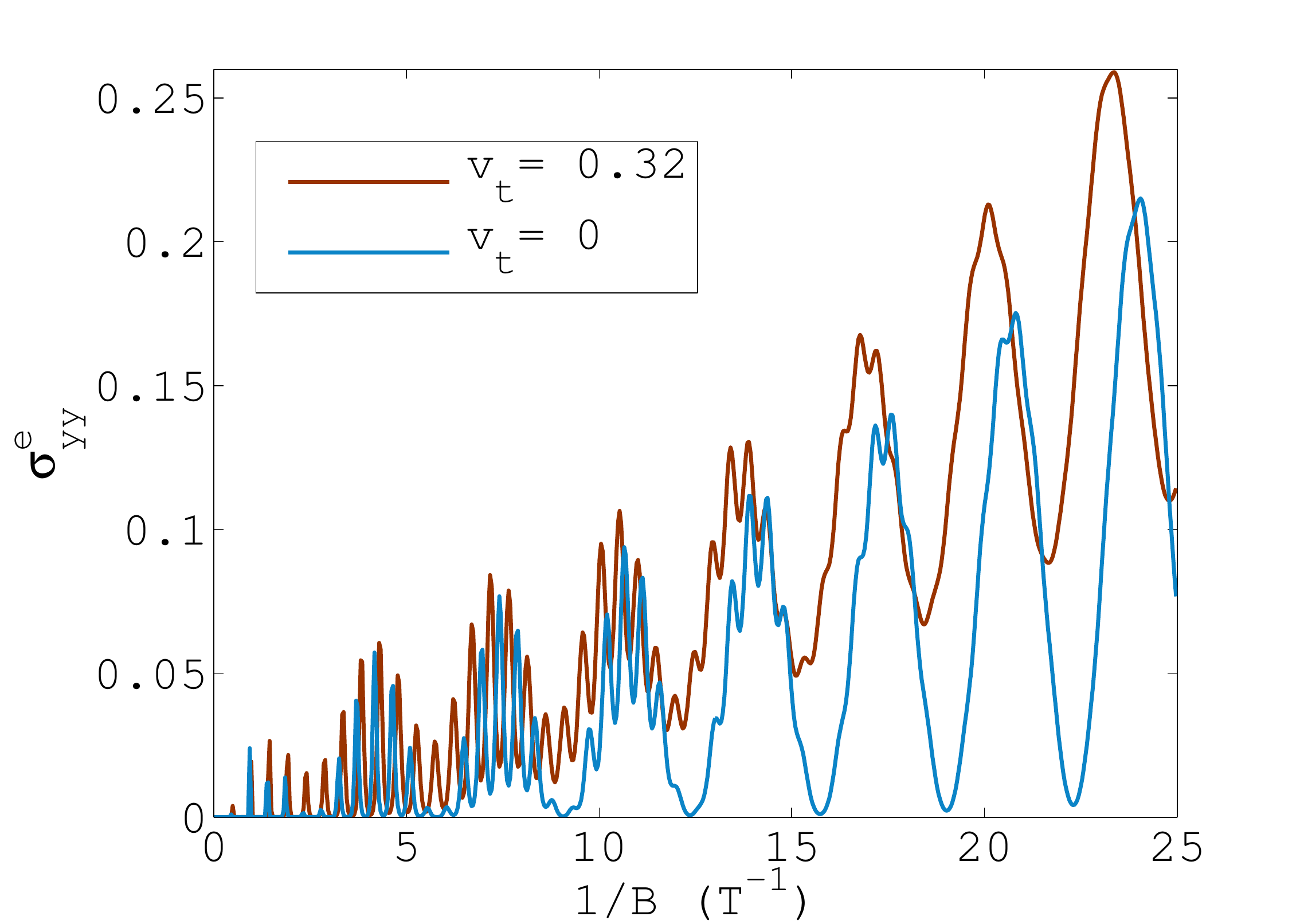}}
 \subfigure[]
  {\includegraphics[width=.49\textwidth,height=5cm]{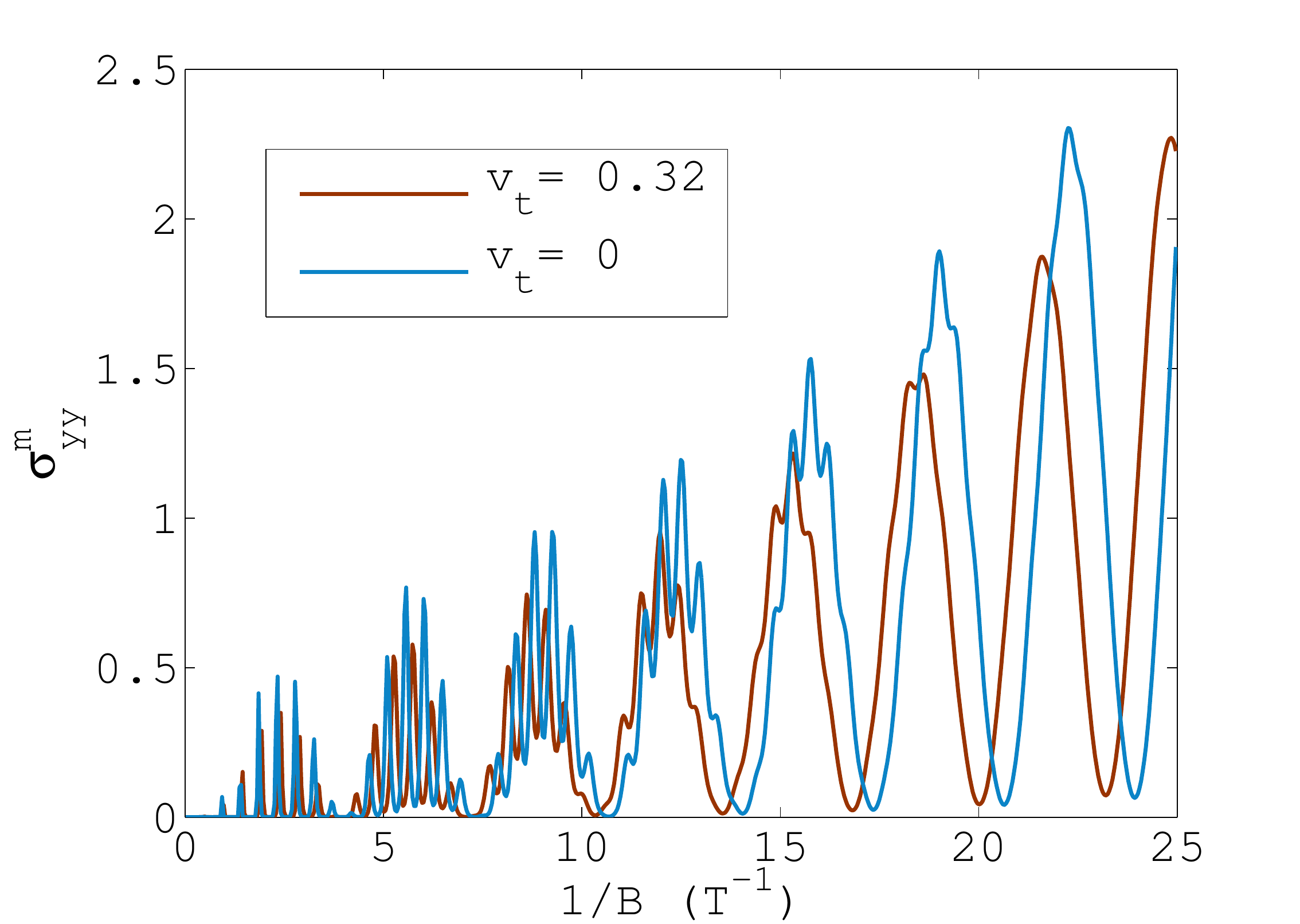}}
\caption{ diffusive conductivity (in units of $e^2/h$) versus inverse magnetic field for (a) electric and (b) magnetic modulation.
The Fermi energy is taken as $0.041$ and $0.035$ eV for $v_t=0$ and $0.32$ in units of $v_0$, respectively.
The modulation period is $a=350$ nm and temperature $T=3$ K. The strength of modulation $V_e=0.5$ meV. The strength of 
the magnetic modulation is taken as $B_m=0.028$ T, such that $V_m^{t}=V_e$.}
\label{weiss_e/m}
\end{figure*}

\subsection{electric modulation}
The application of a weak electric modulation to the borophene sheet is described by the total Hamiltonian
$\mathcal{H}_T^e=\mathcal{H}+V_{e}\sin(\Lambda x)$, where $V_e$ is the modulation strength and $\Lambda=2\pi/a$
with $a$ is the period. Using perturbation theory, we evaluate the first order energy correction as 
\begin{eqnarray}\label{corr_elec}
\Delta E_{\zeta}^{e}&=&\int_0^{L_y}dy\int_{-\infty}^{\infty}\Psi_{\zeta}^{\dagger}({\bf r})V_{e}\sin(\Lambda x)
\Psi_{\zeta}({\bf r})dx\nonumber\\
 &=&\frac{V_{e}}{2}\Big[\xi\beta_b R_{n}(u)\cos(\Lambda \bar{x}_0)\nonumber\\&&-F_{n}(u)\sin(\Lambda \bar{x}_0)\Big].
\end{eqnarray}
 Here $\bar{x}_0=x_0+G_n$ with $G_n=\sqrt{2n\gamma}l_c\beta_b$. Also,
\begin{equation}\label{Fexp}
F_{n}(u)=e^{-u/2}[L_{n-1}(u)+L_{n}(u)]
\end{equation}
and 
\begin{eqnarray}
R_n(u)=\sqrt{\frac{8n}{u}}e^{-u/2}[L_{n-1}(u)-L_{n}(u)],
\end{eqnarray}
where $L_{n}(u)$ is the Laguerre polynomial of order $n$ and $u=\gamma\Lambda^2l_c^2/2$ with $\gamma=(1-\beta_b^2)^{-1/2}$.
The total energy is now $E_{\zeta}^{e}=E_{\zeta}+\Delta E_{\zeta}^{e}$ where $k_y$ degeneracy is now lifted.
The presence of modulation broadens the LLs width by contributing additional energy $\Delta E_{\zeta}^{e}$. The width of the 
LLs broadening \ie band width (in units of $V_e$) is given by $\Delta_e=\sqrt{|\beta_b R_n(u)|^2+|F_n(u)|^2}$, which is oscillatory\cite{PhysRevB.46.4667}
with the inverse magnetic field, as the Laguerre polynomial exhibits oscillatory feature. Note that the first term in Eq.~(\ref{corr_elec}) is 
purely due to the tilting of the Dirac cones which simply vanishes with the tilting parameter $\beta_b=0$. On the other hand,
the second term in the first order energy correction analogous to monolayer graphene case\cite{PhysRevB.75.125429}. 

Now the drift velocity $\mathcal{V}_{\mu(\nu)}$ is obtained as
\begin{equation}
\mathcal{V}_y=-\frac{V_e}{\hbar \Lambda}\frac{u}{\gamma}[\xi\beta_b R_n(u)\sin(\Lambda \bar{x}_0)+F_{n}(u)\cos(\Lambda \bar{x}_0)]
\end{equation}
and $\mathcal{V}_{x}=0$ which suggests that the diffusive conductivity arises along the transverse direction to the applied modulation.
Now, after inserting $\mathcal{V}_y$ into Eq.(\ref{diff}), we obtain diffusive conductivity as
\begin{equation}\label{sigma_elec}
 \sigma_{yy}^{e}=\frac{e^2}{h}\frac{V_e^2}{4\Gamma_0}\frac{u}{\gamma}\sum_{n}
 \left[-\frac{\partial f_{n}}{\partial E}\right]\{[F_{n}(u)]^2+[\beta_b R_n(u)]^2\},
\end{equation}
where $\Gamma_0$ is the impurity-induced broadening. Here, we assume that the collisional time $\tau(E_{\zeta})$ is very insensitive 
to the energy \ie $\tau(E_{\zeta})\simeq \tau_{_0}$ which is a good approximation under low magnetic field regime. We have also substituted 
$\Gamma_0\approx\hbar/\tau_{_0}$. The major effects of modulation arise via non-zero drift velocity. On the other hand,
modulation effect on Fermi distribution function is very small, and hence we ignore it. The diffusive conductivity in the 
above Eq.~(\ref{sigma_elec}) is oscillatory with magnetic field because of the oscillatory nature of the band width ($\Delta_e$).
This oscillation is known as Weiss oscillation. 

For the numerical plots, we use the following physical parameters:
modulation period $a=350$ nm, charge density $n_e=10^{15}m^{-2}$, and temperature $T=3$ K. 
The diffusive conductivity for the electric modulation is plotted with inverse magnetic field in Fig.~(\ref{weiss_e/m})a.
To explore the effects of the tilted Dirac cones, we consider both situations \ie $v_t=0$ and $v_t= 0.32$ unit. The diffusive conductivity
exhibits the Weiss oscillation at low magnetic field regime with the inverse magnetic field. However, with the 
increase of the magnetic field, SdH oscillations start to superimpose over the Weiss oscillation. The SdH oscillations appear
as small oscillation over the envelope of the Weiss oscillation. The tilted Dirac cones cause a significant changes
in the frequency of the Weiss oscillation too. 
To understand the effect of tilting Dirac cones more explicitly, we shall obtain an approximate analytical expression
of the diffusive conductivity. Following the Refs.[\onlinecite{PhysRevB.75.125429,PhysRevB.77.195421}], the diffusive
conductivity for electric modulation can be simplified to an analytical form by using the higher Landau level approximation
\begin{equation}\label{asymp}
e^{-u/2}L_{n}(u)\rightarrow \frac{1}{\sqrt{\pi\sqrt{nu}}}\cos\big(2\sqrt{nu}-\frac{\pi}{4}\big)
\end{equation}
as
\begin{eqnarray}
 \sigma^{e}_{yy}&\simeq& \frac{e^2}{h}\frac{\beta_W}{8\pi^2\Gamma_0\gamma}\Big\{U_0^e-U_1^eR^W\left(\frac{T}{T_W}\right)+2R^WU_1^e\left(\frac{T}{T_W}\right)
 \nonumber\\&&\cos^2\left[2\pi\left(\frac{f}{B}-\frac{1}{8}\right)\right]\Big\}.
\end{eqnarray}
Here,$U_0^e=(V_{e})^2(1+2\beta _b^2)$ and $U_1^e=V_e^2(2\beta_ b^2-1)$. The amplitude of the conductivity is governed by $U_0^e$ which
indicates that it enhances with the tilting of Dirac cones. On the other hand, the Weiss oscillation amplitude is determined by 
the factor $U_1^e$ which is suppressed by the tilting feature of the Dirac cones. The frequency of the Weiss oscillation is
given by $f=E_F\gamma^2/(ev_c a)$. It is clearly seen that the frequency is enhanced by a tilt dependent term $\gamma^{2}=1.20$.
Note that in comparison to graphene, it is not only the tilt parameter which enhances the frequency of the Weiss oscillation,
but also the Fermi velocity ($v_c=0.77\times 10^6$ m/sec) which is smaller than its counterpart in graphene ($v_F=3\times 10^6$m/sec).
Also, $\beta_{W}=(k_BT_W)^{-1}$ with the characteristic temperature $T_{W}=eav_cB/[4(\pi\gamma)^2 k_B]$ which is lowereda by the
tilt parameter. The temperature also induces a damping  to the Weiss oscillation amplitude, which is described by
\begin{equation}
R^W\big(\frac{T}{T_W}\big)=\frac{T/T_W}{\sinh(T/T_W)}.
\end{equation}

\subsection{magnetic modulation:}
Now we consider the case when the perpendicular magnetic field is weakly modulated without any electrical modulation. The underline physics 
of the charge carriers in the presence of a modulated magnetic field is believed to be closely related to {\it composite fermions} in the fractional 
quantum Hall regime\cite{PhysRevB.63.113310}. Under the weak magnetic field and low temperature regime, extensive theoretical works of the 
Weiss oscillation exist from usual $2$DEG to monolayer graphene (as mentioned in the Sec.~\ref{intro}). Along the same line, we investigate Weiss 
oscillation in a magnetically modulated borophene.

First, we evaluate the first order energy correction due to magnetic modulation. Let the perpendicular magnetic field be modulated 
very weakly as ${\bf B}=[B+B_{m}\cos(\Lambda x)]\hat{z}$, where $B_{m}\ll B$, describes the vector potential under the Landau gauge 
${\bf A}=[0,Bx+(B_m/\Lambda)\sin(\Lambda x)]$. Similar to the case of electric modulation, the total Hamiltonian can now be split 
into two parts as $\mathcal{H}_T^m=\mathcal{H}+\mathcal{H}_m$, where $\mathcal{H}$ is the unperturbed Hamiltonian
and $\mathcal{H}_{m}$ is the modulation induced perturbation which can be written as
\begin{equation}
 \mathcal{H}_{m}=\xi\frac{eB_m\sin(\Lambda x)}{\Lambda}\left(\sigma_0v_t+\sigma_yv_c\right).
\end{equation}
Using the unperturbed wave function, the first order energy correction due to the magnetic modulation $\mathcal{H}_{m}$ is evaluated as
\begin{equation}\label{corr_mag}
\Delta E_{\zeta}^{m}=\frac{1}{2}[\xi V_m^{ct}F_n(u)\sin(\Lambda \bar{x}_0)+V_m^{tc}R_n(u)\cos(\Lambda \bar{x}_0)].
\end{equation}
Here, $V_{m}^{ct}=(\beta_b V_m^c-V_m^t)$ and $V_{m}^{tc}=(\beta_b V_m^t-V_m^c)$ with $V_{m}^{t}=eB_m v_t/\Lambda$ 
and $V_{m}^{c}=eB_m v_c/\Lambda$. In the above energy correction, in Eq.~(\ref{corr_mag}), the 
terms involving $V_m^t$ and $\beta_b$ are purely due to the tilting feature of the Dirac cones. The above equation can be 
reduced to the case of magnetically modulated graphene~\cite{PhysRevB.77.195421} by setting $\beta_b=V_m^t=0$. The width of the
LLs broadening is $\Delta_m=\sqrt{[V_m^{ct}F_n(u)]^2+[V_m^{tc}R_{n}(u)]^2}$.
Now the group velocity is found to be 
\begin{equation}
\mathcal{V}_y=\frac{u}{\hbar\gamma \Lambda}[\xi V_m^{ct}F_n(u)\cos(\Lambda \bar{x}_0)-V_m^{tc}R_n(u)\sin(\Lambda \bar{x}_0)]
\end{equation}
and $\mathcal{V}_{x}=0$. Following the same procedure, as in the electric modulation case, we obtain the diffusive conductivity as
\begin{equation}\label{sigma_mag}
\sigma_{yy}^{m}=\frac{e^2}{h}\frac{u}{4\gamma\Gamma_0}\sum_{n}\left[-\frac{\partial f_{n}}{\partial E}\right]
\{\left[V_m^{ct}F_{n}(u)\right]^2+\left[V_m^{tc}R_{n}(u)\right]^2\}.
\end{equation}
Note that the diffusive conductivity for electrically [Eq.~(\ref{sigma_elec})] and magnetically modulated [Eq.~(\ref{sigma_mag})] borophene
are independent of the valley index. The first term inside the third bracket of the above equation arises due to the tilting of the 
Dirac cones, and gives extra contribution to the diffusive conductivity. Following the similar approach as in the electric modulation case,
we obtain the analytical expression of diffusive conductivity as
\begin{eqnarray}\label{ana_mag}
 \sigma^{m}_{yy}&\simeq& \frac{e^2}{h} \frac{\beta_WU_0}{8\pi^2 \gamma\Gamma_0}\Big\{1-\left(\frac{U_{1}}{U_0}\right)R^W\left(\frac{T}{T_W}\right)
 +\nonumber\\&&2\left(\frac{U_{1}}{U_0}\right)R^W\left(\frac{T}{T_W}\right)\sin^2\left[2\pi\left(\frac{f}{B}-\frac{1}{8}\right)\right]\Big\}
\end{eqnarray}
with $U_0=(V_{m}^{ct})^2+(\sqrt{2}V_m^{tc})^2$ and $U_1=(\sqrt{2}V_{m}^{tc})^2-(V_m^{ct})^2$.  The amplitude of the Weiss 
oscillations is governed by the factor $U_1$ and it is suppressed in presence of the tilt induced term $V_m^{t}$.
The tilt induced additional contribution to the diffusive conductivity does not appear in non-tilted Dirac material
graphene. 

The diffusive conductivity for magnetic modulation is plotted numerically
in Fig.~(\ref{weiss_e/m})b, where Weiss oscillation is found to be weakly suppressed. The origin of this suppression can be
understood from the analytical expression of diffusive conductivity given in Eq.~(\ref{ana_mag}). The presence of the tilt induced
term $V_{m}^{t}$ reduces the amplitude of the oscillation, governed by $U_1$.  Note that we have taken the strength of magnetic
modulation $B_m=0.028$ T, such that $V_e=V_{m}^{t}=0.5$ meV and $V_{m}^{c}=1.2$ meV. Note that, in comparison to the case
of electrical modulation, the amplitude of Weiss oscillation is enhanced.
\begin{figure}[thpb]
\centering
{\includegraphics[width=.95\linewidth,height=7cm]{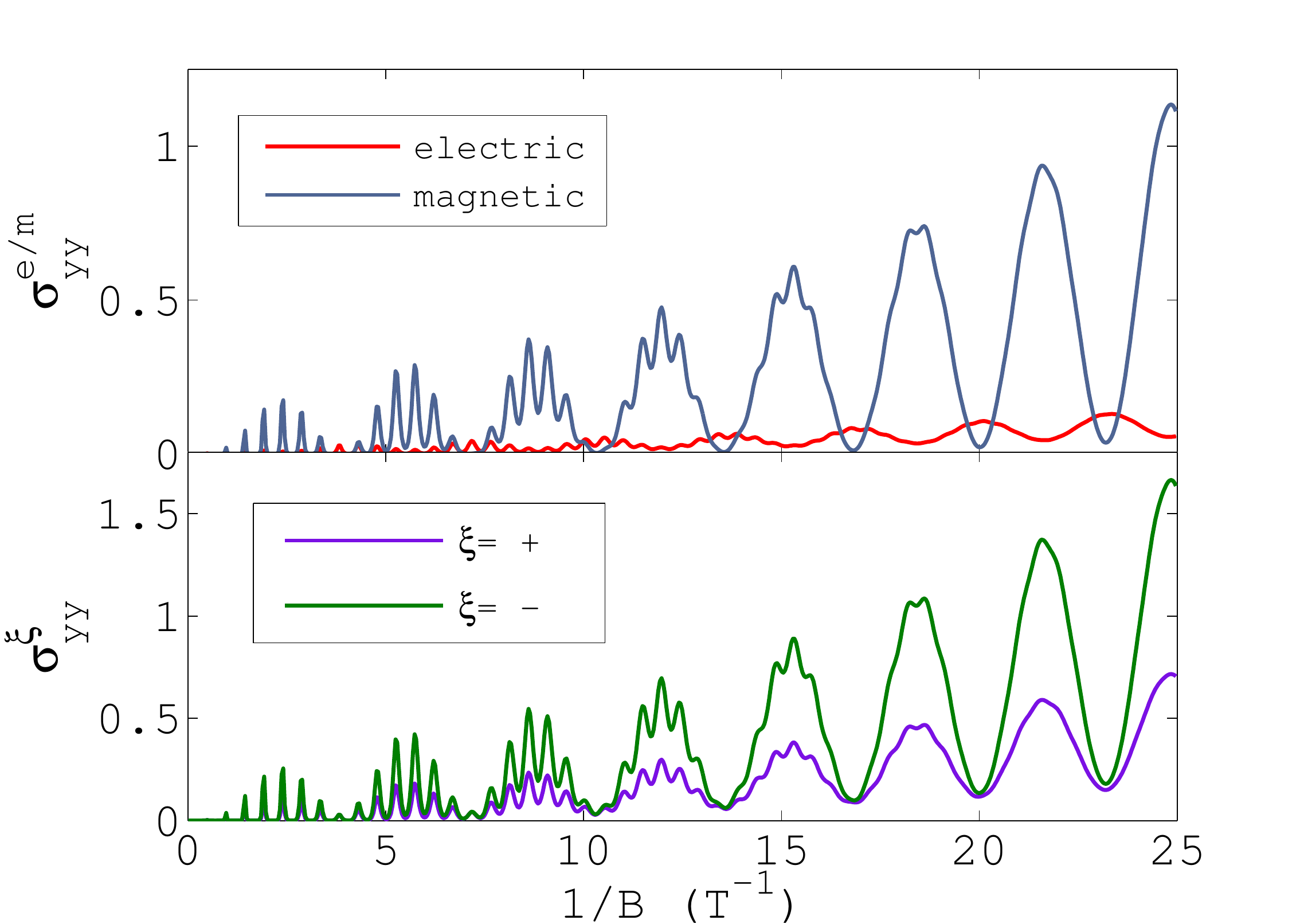}}
\caption{Diffusive conductivity (in units of $e^2/h$) versus inverse magnetic field for
electric and magnetic modulation (upper panel) and in presence of both out-of phase
modulations (lower panel) in each valley. The tilt velocity $v_t=0.32v_0$ and
Fermi energy is $0.035$ eV. All other parameters are taken same as in Fig.~(\ref{weiss_e/m}).}
\label{out_phase}
\end{figure}
\subsection{presence of both modulation}
Now we consider the situation when both types of modulation \ie electric and magnetic are present together. The presence of both modulations 
may give rise some new features to the Weiss oscillation. In usual $2$DEG\cite{PhysRevB.47.1466}, it was found that Weiss oscillation can be
pronounced or suppressed depending on whether both modulations are in phase or out of phase. Recently, we have observed that the 
presence of both modulations can break particle-hole symmetry in Dirac materials like graphene and $\alpha-T_3$ lattice\cite{PhysRevB.96.045418}.
\begin{figure}[thpb]
\centering
{\includegraphics[width=.95\linewidth,height=5cm]{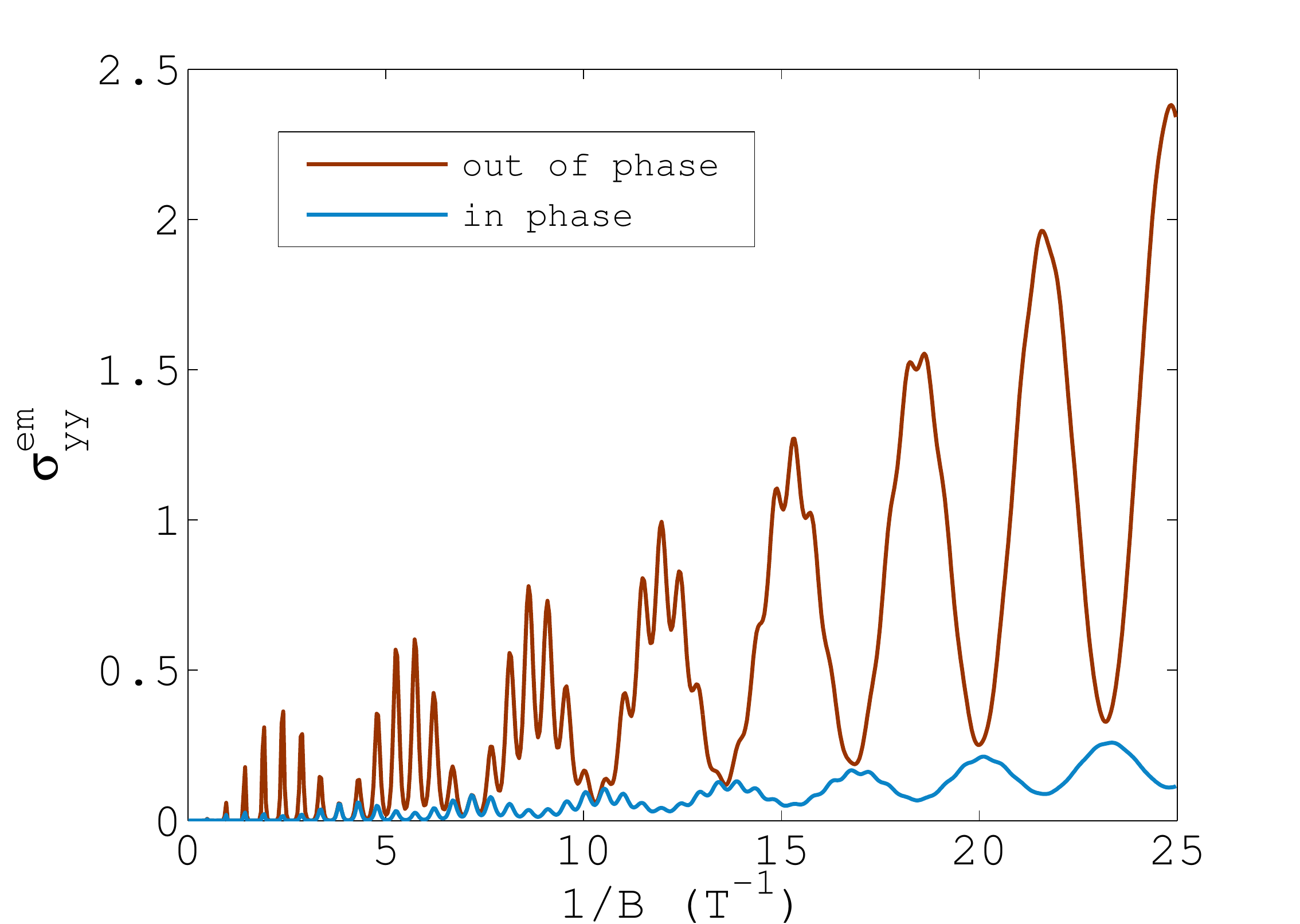}}
\caption{Diffusive conductivity (in units of $e^2/h$) versus inverse magnetic field in presence of the both modulations.
All other parameters are kept same as in the Fig.~(\ref{out_phase}).}
\label{both_modu}
\end{figure}
First we consider that the electric and magnetic modulations are in out-of phase \ie $V_e\sin(\Lambda x)$ and $B_m\cos(\Lambda x)$, respectively.
The total first order energy correction is evaluated to be
\begin{equation}
 \Delta E^{em}_{\zeta}=\frac{1}{2}\left[W_1^{\xi}R_{n}(u)\cos(\Lambda \bar{x}_0)+W_2^{\xi}F_n(u)\sin(\Lambda \bar{x}_0)\right],
\end{equation}
where $W_1^{\xi}=(\xi V_e\beta_b+V_m^{tc})$ and $W_2^{\xi}=(\xi V_m^{ct}-V_e)$.
The group velocity of charge carriers under the combined effects of out of phase both modulations is evaluated as
\begin{equation}
 \mathcal{V}_y=\frac{u}{\hbar \gamma\Lambda}\left[W_2^{\xi}F_n(u)\cos(\Lambda \bar{x}_0)-W_1^{\xi}R_{n}(u)\sin(\Lambda \bar{x}_0)\right]
\end{equation}
and $\mathcal{V}_x=0$. Following the same procedure as in the case of electric/magnetic modulation,
we obtain the valley dependent diffusive conductivity as 
\begin{equation}
\sigma_{yy}^{\xi}\simeq\frac{e^2}{h}\frac{\beta_W u}{8\gamma\Gamma_0}\sum_{n}\left[-\frac{\partial f_{n}}{\partial E}\right]
\{[W_1^{\xi} R_n(u)]^2+[W_2^{\xi} F_{n}(u)]^2\}.
\label{elec_mag}
\end{equation}
The most remarkable point here is that now the diffusive conductivity is very sensitive to the valley index
and can cause sizable valley polarization, which is attributed to the presence of the tilt induced term $V_{m}^t$ and $\beta_b$.
In non-tilted Dirac cones like graphene, valley polarization does not appear even in presence of the both modulations because
of the absence of the term $V_m^{t}$ and $\beta_b$. Using the similar approach as in the electric/magnetic modulation, analytical
form of the diffusive conductivity can be obtained as
\begin{eqnarray}\label{ana_both1}
 \sigma^{\xi}_{yy}&=& \frac{e^2}{h} \frac{\beta_WU_0^{\xi}}{16\pi^2 \gamma\Gamma_0}\Big\{1-\left(\frac{U_{1}^{\xi}}{U_0^{\xi}}\right)
 R^W\left(\frac{T}{T_W}\right)\nonumber\\&&+2\left(\frac{U_{1}^{\xi}}{U_0^{\xi}}\right)R^W\left(\frac{T}{T_W}\right)
\sin^2\left[2\pi\left(\frac{f}{B}-\frac{1}{8}\right)\right]\Big\},\nonumber\\
\end{eqnarray}
where $U_0^{\xi}=(W_1^{\xi})^2+(W_{2}^{\xi})^2$ and $U_1^{\xi}=(W_1^{\xi})^2-(W_{2}^{\xi})^2$. 
The amplitude of the Weiss oscillation is determined by a valley dependent factor $U_{1}^{\xi}$.
In $K$-valley ($\xi=+$), the Weiss oscillation amplitude is much suppressed than $K'$-valley ($\xi=-$),
which is shown in the lower panel of Fig.~(\ref{out_phase}).
The upper panel of this figure shows that when electric and magnetic modulations are applied individually, the
Weiss oscillation in one valley ($K$ or $K'$) exhibits opposite phase with an amplitude mismatch. The origin of 
the opposite phase is well addressed in Ref.[\onlinecite{PhysRevB.77.195421}]. However, when both modulations are
applied together then two valleys respond differently.
The Weiss oscillation amplitudes are enhanced in both valleys but the enhancement in $K'$-valley is much higher than 
$K$-valley, as shown in the lower panel of Fig.~(\ref{out_phase}). These features can be understood from analytical
expression of diffusive conductivity in Eq.~(\ref{ana_both1}). In $K$-valley, the amplitude of Weiss
oscillation is determined by $U_{1}^{+}$ which is smaller than its counterpart in $K'$-valley \ie $U_{1}^{-}$.

On the other hand, if we consider that the both modulations are in the same phase \ie electric modulation is
$V_e\cos(\Lambda x)$ and the magnetic modulation is of the form of $B_m\cos(\Lambda x)$, then diffusive conductivity will be 
\begin{equation}
\sigma_{yy}^{em}=\frac{e^2}{h}\frac{u}{4\gamma\Gamma_0}\sum_{n}\left[-\frac{\partial f_{n}}{\partial E}\right]\{[W_{3}(n,u)]^2+[W_{4}(n,u)]^2\},
\label{elec_mag2}
\end{equation}
where $W_{3}(n,u)=[V_e\beta_b R_n(u)+V_m^{ct}F_n(u)]$ and $W_{4}(n,u)=[V_m^{ct} R_n(u)-V_eF_n(u)]$.
In this case, diffusive conductivity does not depend on the valley index. A similar analytical expression can also
be found by following the same approach. The Weiss oscillation for the presence of in-phase and out-of phase both
modulations are presented together in Fig.~(\ref{both_modu}), which shows that the Weiss oscillations in both cases
are in opposite phase with amplitude mismatch. This features are the direct consequences of 
the total effective energy correction in both cases.
\begin{figure*}
 \subfigure[]
 {\includegraphics[width=.49\textwidth,height=5cm]{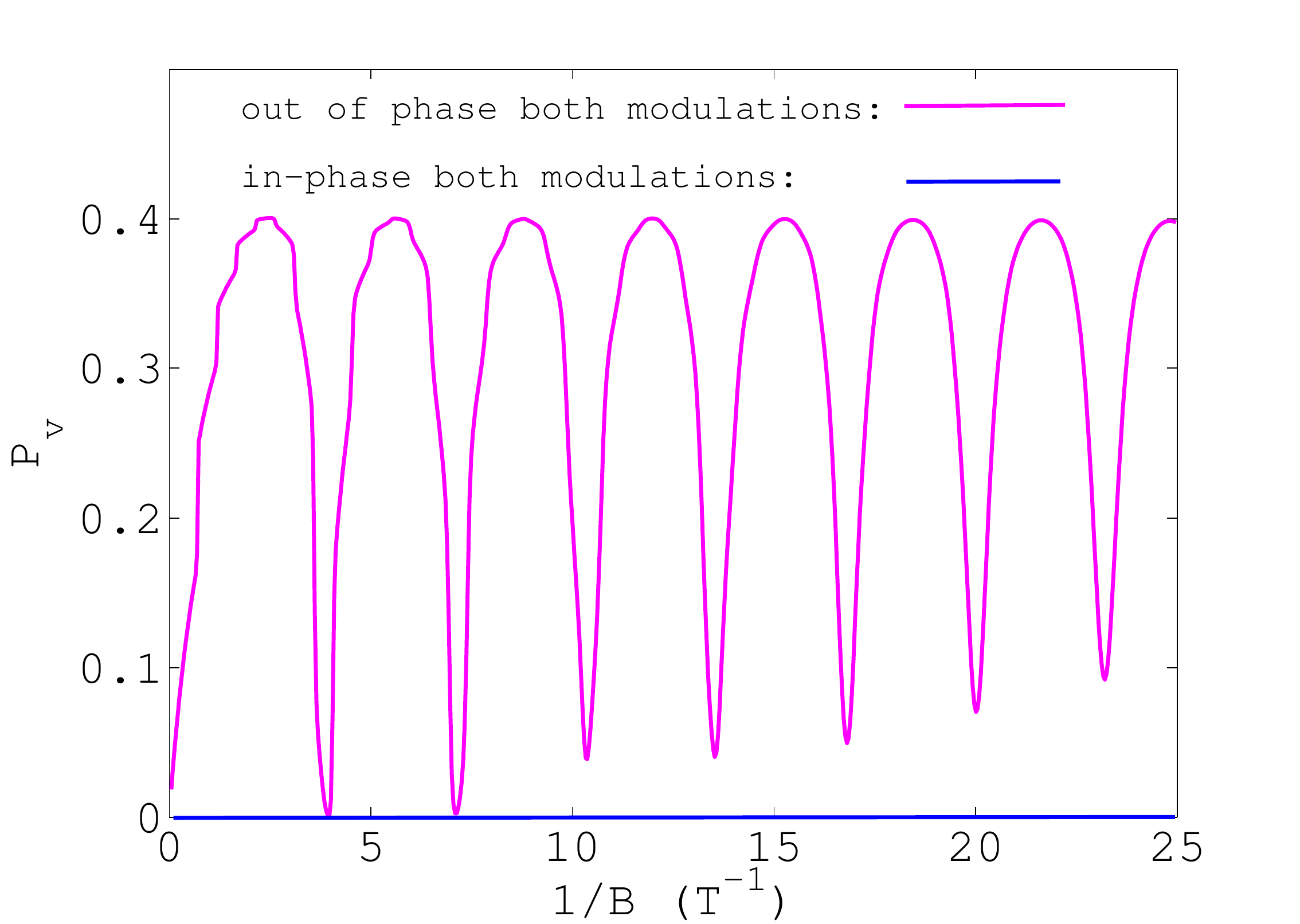}}
 \subfigure[]
 {\includegraphics[width=.49\textwidth,height=5cm]{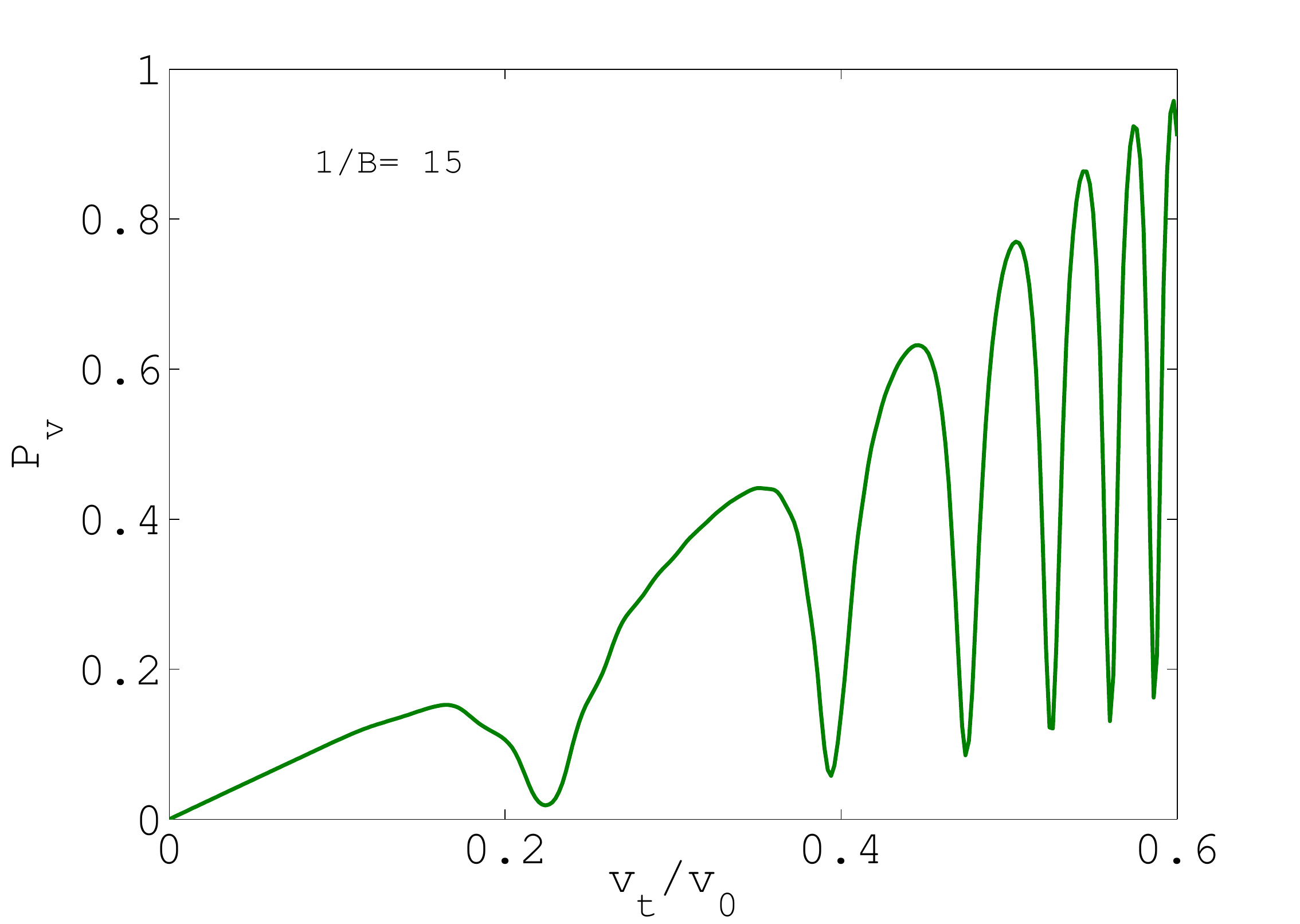}}
\caption{Valley polarization versus (a) inverse magnetic field  and (b) tilt velocity ($v_t$). Fermi energy is 
kept fixed at $E_{F}=0.035$eV. All parameters are kept same as Fig.~(\ref{weiss_e/m}).}
\label{pola}
\end{figure*}
As we have seen that the diffusive conductivity may be sensitive to valley index depending on the phase relationship between 
electric and magnetic modulations, it is interesting to examine the valley polarization, for which we plot it versus inverse 
magnetic field in Fig.~(\ref{pola}).
Because of the unequal suppression of Weiss oscillations in two valleys in presence of the out-of phase both modulations,
a sizable valley polarization arises in the diffusive conductivity. To plot valley polarization, we define it as 
\begin{equation}\label{pola_defn}
P_v=\frac{\sigma_{yy}^{+}-\sigma_{yy}^{-}}{\sigma_{yy}^{+}+\sigma_{yy}^{-}}.
\end{equation}
The valley polarization oscillates with the inverse magnetic field with the frequency of Weiss oscillation as shown in Fig.~(\ref{pola})a. 
Appearance of valley polarization strongly depends on the phase relationship between both modulations.
Valley polarization appears in Weiss oscillation only when electric and magnetic modulations are in out of phase. It is also
intereting to examine the evolution of valley polarization with $v_t$. In Fig.~(\ref{pola})b, we also show evolution of valley polarization
in diffusive conductivity with the smooth variation of tilt velocity $v_t$. It shows that valley polarization oscillates with $v_t$,
which can be easily understood from the fact that Weiss oscillation frequency has a strong dependency on $v_t$ too. The valley
polarization shows emergence of regular peaks with increasing height towards $1$ with the increase of $v_t$. 
We mention here that we have taken Fermi energy ($0.035$ eV) as constant while ploting the Fig.~(\ref{pola})b although a weak
dependecy of Fermi energy on $v_t$ exists as shown in Fig.~(\ref{Fermi}). The rise of valley polarized Weiss oscillation in diffusive conductivity
is one of our main results which differs from graphene. Here, we mention that the valley polarized Weiss oscillation was predicted
in electrically modulated silicene\cite{PhysRevB.90.125444} too. However, in that case a gate voltage between two planes of
sub-lattices is necessary in addition to the strong spin-orbit interaction.

Finally, we discuss if tilt parameter can be extracted from the Weiss oscillation experiment. The 
frequency of the Weiss oscillation can be easily obtained from magnetoresistance measurement of borophene,
which can be directly used to extract the tilt parameter once we know the Fermi level and Fermi velocity.
The direct method of obtaining Fermi level and velocity was recently reported in Ref.[\onlinecite{PhysRevLett.108.116404}].
%

\section{Summary and conclusions}\label{sec4}
In this work, we have studied magnetotransport properties of a 2D sheet of the polymorph of a periodically modulated
$8$-$Pmmn$ borophene which exhibits tilted anisotropic Dirac cones. We have evaluated the modulation induced diffusive
conductivity by using the linear response theory. The diffusive conductivity exhibits the Weiss oscillation with the inverse magnetic field, the frequency
of which is enhanced by the tilted feature of the Dirac cones. The amplitude of Weiss oscillation is also enhanced or suppressed
depending on the types of modulation. Most remarkably, we have found that the presence of out-of phase electric and magnetic modulations can 
cause very high valley polarization in Weiss oscillation at low magnetic field. The appearance of the valley polarization in the Weiss 
oscillation is the direct manifestation of the tilted Dirac cones in borophene. It is in complete contrast to the 
non-tilted isotropic Dirac material-graphene where such valley polarization does not appear. 

As far as the practical realization of this material is concerned, a borophene structure can be formed on the surface of 
Ag(111) as reported recently in Ref.[\onlinecite{PhysRevLett.118.096401}]. On the other hand, periodic modulation can be imparted
to the system by several methods. For example, an array of biased metallic strips on the surface of a $2$D electronic system has
been used by Winkler \etal \cite{PhysRevLett.62.1177} to achieve electric modulation. Magnetic modulation can be generated by placing a few patterned
ferromagnets or a superconductor on the surface of the $2$D material~\cite{izawa1995,PhysRevLett.74.3009,PhysRevLett.74.3013}. 
\begin{acknowledgements}
SFI thanks to Tarun K. Ghosh and Tutul Biswas for useful comments. Authors also thank the anonymous referee for 
pointing out an error in the calculation. One of us AMJ thanks DST, India for J.C. Bose National  Fellowship.
\end{acknowledgements}
\bibliography{bibfile}{}
\end{document}